\documentclass[]{spie}  

\usepackage{amsmath,amsfonts,amssymb}
\usepackage{graphicx}
\usepackage[colorlinks=true, allcolors=blue]{hyperref}
\usepackage{comment}
\usepackage[T1]{fontenc}
\usepackage[utf8]{inputenc}

\title{Simulation and development of astrophotonic H-band nulling beam combiner for the CHARA Array}

\author[a]{Samantha G. Morrison}
\author[a]{Pradip Gatkine}
\author[a]{Michael P. Fitzgerald}
\affil[a]{Department of Physics and Astronomy, University of California, Los Angeles, 475 Portola Plaza, Los Angeles, USA}

\authorinfo{Further author information: (Send correspondence to S.G.M.)\\S.G.M.: E-mail: smorrison20@ucla.edu}

\begin{document} 
\maketitle

\begin{abstract}
Nulling interferometry is a promising method for direct detection and characterization of faint stellar companions such as exoplanets and faint binary sources. The dominant constraint on instrument performance is instability noise from vibrations, thermal expansions, and other systematic sources. 
Photonic instruments have footprints of a few centimeters and enable advanced light control architectures. Astrophotonic implementations can include active phase control via Mach-Zehnder interferometers (MZIs) to actively minimize residual optical path errors at the location of beam combination. The compact size of photonic chips also makes them a promising solution for space-based applications such as the upcoming Habitable Worlds Observatory.
The Center for High Angular Resolution Astronomy (CHARA) Array (Mt. Wilson, CA) is equipped with the world’s longest baseline in the near-infrared at 330 meters long. CHARA therefore has the most sensitive angular resolution in the near-infrared. A nuller at CHARA would have access to a novel parameter space essential for exoplanet discovery and characterization.
The first step in preparing a nulling beam combiner for CHARA is to develop compelling, realistic, and well-defined science cases. Simulation suites like GENIEsim\cite{absil_geniesim_2003}, SCIFYsim\cite{laugier_asgardnott_2023}, and PHRINGE/LIFEsimMC\cite{huber_robust_2025} quantify the technical requirements to reach such science goals in their respective complex systems. The CHARA Array Response Model (CHARM) is the first-ever simulator built specifically for CHARA, opening up new pathways for instrument simulation. CHARM is a flexible simulation tool of an H-band, self-calibrated astrophotonic nulling beam combiner with active phase control for the CHARA Array.
In this paper, we present an update on the development of our photonic nulling beam combiner for CHARA. We include early results from CHARM, demonstrating that the photonic nuller has the potential to detect and characterize a wide range of bright stellar companion systems.
 
\end{abstract}


\section{INTRODUCTION}
\label{sec:intro}  

Almost 6,300 exoplanets have been confirmed (as of June 2026)\cite{christiansen_nasa_2025}. The majority of planet discoveries to date have occurred via indirect methods, the most prolific of which are the radial velocity and transit methods. Direct detection methods, including coronography and nulling interferometry, have lagged due to technological challenges. However, direct observations offer distinct advantages: direct observation can break important degeneracies in parameter space by providing true spectral, size, orientation, distance, or multiplicity measurements. There is much interest in developing direct detection methods that can observe closer in, fainter systems, with better contrast, and other improvements. Upcoming missions such as NASA's Habitable Worlds Observatory \cite{national_academies_of_sciences_pathways_2023} and ESO's Large Interferometer For Exoplanets (LIFE) hope to make use of these technologies\cite{quanz_large_2022}. If we hope to have them ready in time, we need to take the first steps in developing the novel technologies now. Early steps include laboratory testing of individual components, installation at ground-based observatories, space qualification, and simulation work. 

Nulling interferometry has unique abilities among direct detection techniques. Unlike coronagraphy, nulling is not necessarily limited by an inner-working angle (IWA), and is thus better-suited to observe close-in companions. Of particular interest are exoplanets orbiting within the snowline of their host stars, some of which may be within the habitable zone of their host stars. The development of nulling interferometry as a technique is therefore a crucial stepping stone on the path toward upcoming NASA and ESA missions which aim to discover Earth analogs: rocky planets orbiting within the habitable zone. These missions include: HWO, LIFE, PLATO, Roman, PANDORA, and Ariel. 

Existing nulling interferometers are limited by the stability of their noise floors. Vibrations and thermal expansions of the instrument can introduce unwanted optical path delays, and differential piston can distort the phase correction. Such perturbations degrade the interferometric null depth and can mimic an exoplanet signal. Kernel nulling addresses this issue by exploiting the inherent symmetries of an interferometric system. Subtracting neighboring sets of mirrored-phase outputs (``enantiomorphs'') produces observables that are robust against residual phase orders to the second order \cite{martinache_kernel_2010}. In the case of an optical path difference (OPD) error, the two enantiomorphs receive the same excess photons, so their subtraction is immune to the error.

The Center for High Angular Resolution Astronomy (CHARA) is an interferometer located at Mt. Wilson, CA, and operated by Georgia State University. CHARA consists of six, one-meter telescopes arranged in a Y-shape, with a maximum baseline of 331 meters \cite{brummelaar_first_2005}. This is the longest operational baseline in the optical/near infrared anywhere in the world. Interferometric resolution is defined as $R= \lambda/2B$, where $B$ is the baseline length. Meanwhile, single-aperture resolution is defined by the diameter of the aperture: $R = \lambda/D$. As typical baseline lengths greatly exceed aperture sizes, interferometric observations can resolve much smaller angular resolutions than single-aperture observations. Additionally, longer baselines lead to more sensitive resolution, which gives CHARA the most sensitive angular resolution of any operational system in the H-band, ground- or space-based. 

The H-band ($\lambda \sim 1.65 \mu \mathrm{m}$) is important for exoplanet science cases. Hot planets like Hot Jupiters emit significantly at these wavelengths. Their parent stars are sometimes cool enough for the contrast ratios to be advantageous. H-band technology also benefits from the long history of telecommunications industry development, which has focused on these wavelengths. More components are off-the-shelf ready, bringing down prices and raising repeatability.

Photonic circuits are quite like electronic circuits except that they propagate photons rather than electrons. Photonic integrated circuits (PICs) can support advanced architectures, including three-dimensional waveguide crossings, self-configuring active feedback loops, or complex combination schemes. Real-time correction via active photonic control is possible using Mach-Zehnder interferometer (MZI) meshes, which apply heat to a small patch of waveguide to correct phase aberrations \cite{sercel_thermal_2024}. These features are necessary for advanced techniques like nulling interferometry, which requires extreme precision and stability. Important photonic building blocks for nulling include the double Bracewell architecture, self-calibrating MZI meshes, and kernel-nulling architectures. PICs are also much smaller than traditional bulk optical systems, which makes them easier to stabilize. When PICs are fed by single-mode fibers, they remain small even as telescope sizes increase. This is a distinct advantage in the age of Extremely Large Telescope (ELT) development.

Simulations are an effective and inexpensive method for testing specific engineering choices. Simulations can define science cases for a newly proposed instrument by quantifying important observables. In the case of a nulling interferometer, such observables include contrast, null depth, separation, and throughput.

\section{METHODS}

To explore the parameter space of a nulling beam combiner at CHARA, we have created the first dedicated end-to-end simulator of the CHARA Array. The CHARA Array Response Model (CHARM; Morrison et al., in prep.) is designed to replicate an observation at CHARA. CHARM can therefore answer questions about the outcome of specific proposed instrument upgrades, including the H-band nulling interferometric chip proposed here.

A few highlighted inputs to the simulator are amplitude and phase stability. Lay (2004) shows that a general aberration can be broken down into its phase, amplitude, and polarization components\cite{lay_systematic_2004}. These aberrations can be further broken down by mode profile. The chip injection system at CHARA involves single-mode fibers, so only the fundamental mode is coupled into the chip's waveguides. The aberrations can therefore be completely described by a single number each. 

For amplitude aberrations, we choose a stability of 0.01\%. For phase aberrations, we choose 3 nm $\Delta$OPD. Current estimates at CHARA expect an amplitude mismatch of approximately 1\% and a residual $\Delta$OPD of around 100 nm \cite{anugu_chara_2026}. By relying on active phase control, we seek to lower performance residuals. This paper represents the expectations of such an improved instrument.

\begin{figure}
    \centering
    \includegraphics[width=0.7\linewidth]{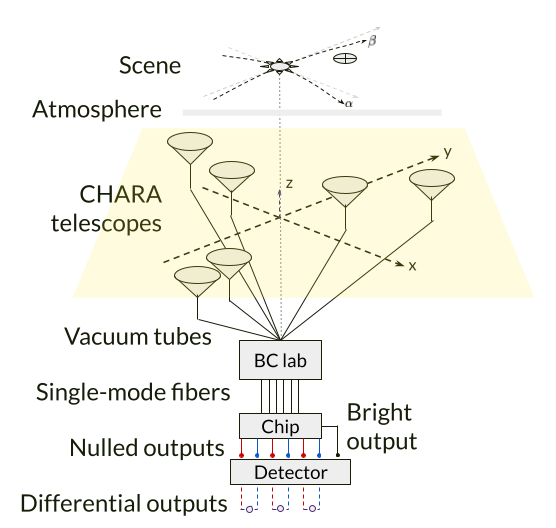}
    \caption{Schematic diagram of the simulated instrument, including coordinate system. The CHARA telescope system is outfitted with the photonic nulling beam combiner detailed here. BC lab = beam combination lab.}
    \label{fig:instrument_schematic}
\end{figure}

The heart of any photonic interferometric instrument is a beam combination chip. The chip contains waveguides that funnel the beams from each telescope through the preferred configuration, with directional couplers (evanescent coupling regions) that permit the different telescopes to interact in a controlled environment. Such beam combination chips can support advanced beam combination architectures with great precision in splitting ratios. A general beam combination chip can be described by a matrix. Here, we simulate a 6-T kernel nulling beam combiner proposed by Laugier et al. (2020) \cite{laugier_kernel_2020}. 

Sensitive nulling on long baselines such as those at CHARA requires extremely precise pathlength matching. To remove residual optical path errors, CHARA’s existing adaptive optics and fringe tracking facilities will be supplemented by active phase control provided by the on-chip MZIs. The resulting noise reduction stands to drastically improve the phase stability of current beam combiners at CHARA. Preliminary laboratory results of predecessor astrophotonic chips show that the phase control and amplitude stability are sufficient to maintain the required null depths throughout an observation.

\begin{figure}
    \centering
    \includegraphics[width=0.9\linewidth]{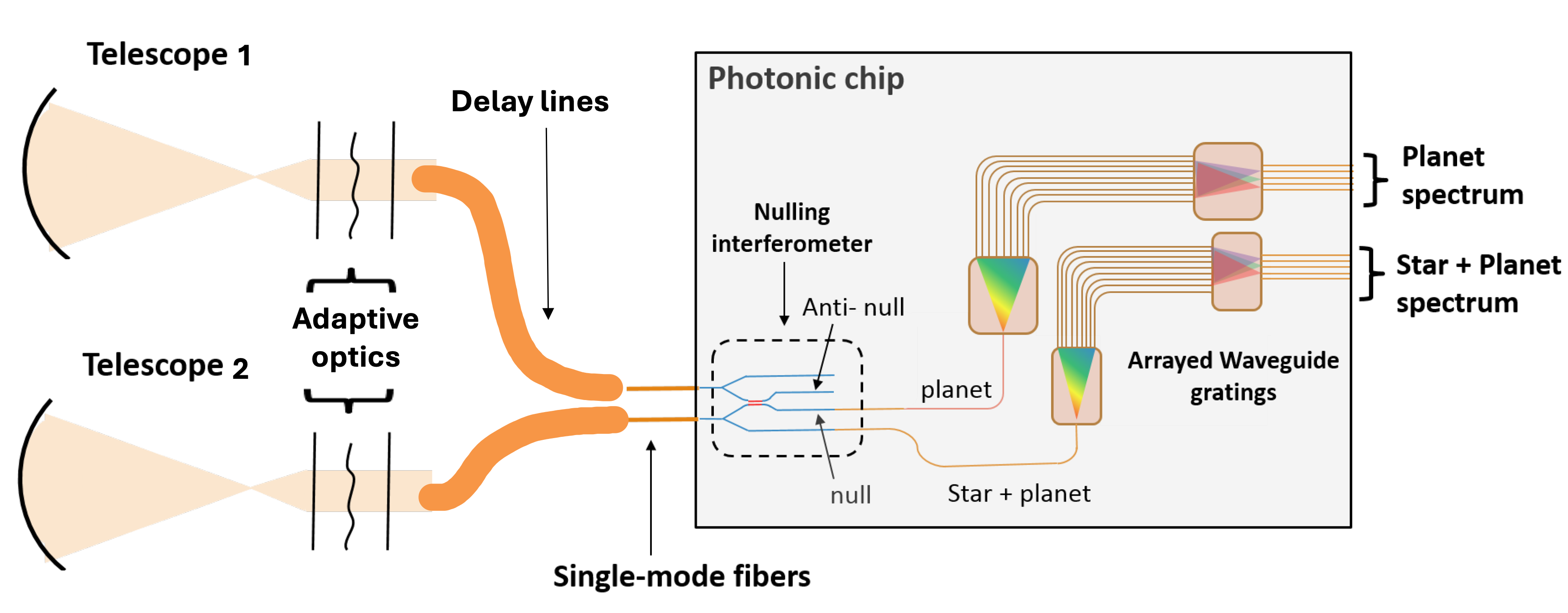}
    \caption{Toy model of a 2-telescope photonic nulling interferometer, including spectroscopic dispersion of the outputs. Adapted from Gatkine et al. (2019)\cite{gatkine_astrophotonic_2019}.}
    \label{fig:2T_toy_model}
\end{figure}

\begin{figure}
    \centering
    \includegraphics[width=0.9\linewidth]{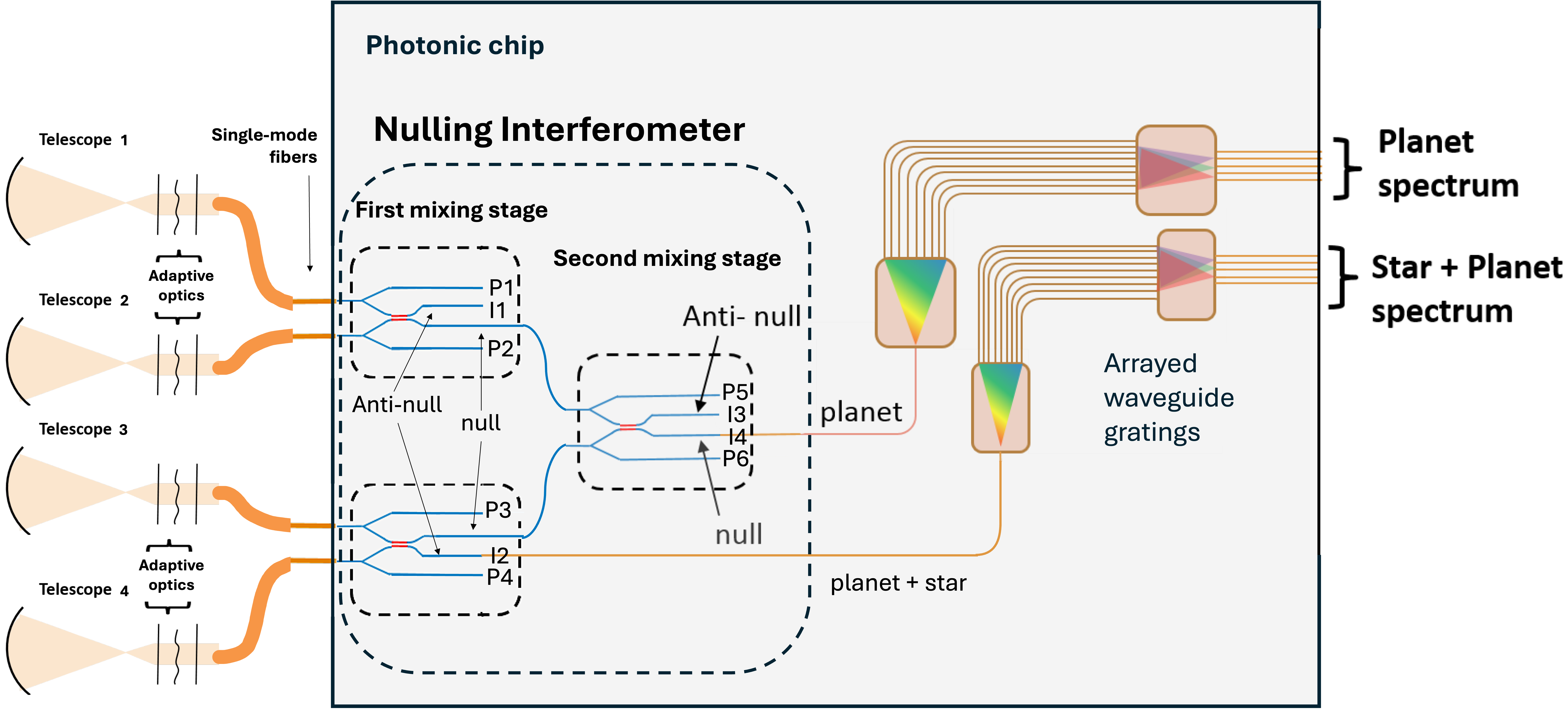}
    \caption{Toy model of a 4-telescope double Bracewell photonic nulling interferometer, including spectroscopic dispersion of the outputs. Outputs P1-P6 are photonic taps, used for monitoring the intensity and/or phase on the science channel. Outputs I1-I4 are the science channels. Adapted from Gatkine et al. (2019)\cite{gatkine_astrophotonic_2019}.}
    \label{fig:4T_toy_model}
\end{figure}

Here, we include a preliminary design sketch for the first design stages of our instrument. While the full instrument will incorporate all 6 CHARA telescopes, the first chips to print will have 2 and then 4 telescope inputs to perform quality checks and proof of concept. These chips will be tested on-sky to inform further instrument development. Figs \ref{fig:2T_toy_model} and \ref{fig:4T_toy_model} show a toy model of a 2-T and 4-T system design, respectively. 

A 2-T system has two inputs (T1, T2) and two science outputs (null, anti-null). There are also photonic taps (P1, P2) that monitor the conditions (amplitude, phase) on each of the incoming single-mode fibers. These taps can be used to reconfigure the coupling element (red crossing) to optimize the splitting ratio in real time. This correction is referred to as active control.

The observing method of a nulling interferometer merits further discussion. Unlike traditional photometric imaging, interferometry does not produce ``pictures'' of a source. Instead, the planetary signal is revealed by a differential intensity modulation in neighboring channels. As the sky rotates overhead, the projected telescope baselines shift. Therefore, an off-axis source passes through the transmission map, creating a distinctive signal. Eq. \ref{eq:SNR} summarizes how to compute a signal-to-noise ratio (SNR) from nulling data. The planet can also be detected by a power spectrum; see Lay (2004) for details \cite{lay_systematic_2004}. 

\begin{equation}
    \text{SNR} = \frac{\text{signal with planet}}{\text{signal without planet}} = 
\frac{\text{standard deviation of planetary signal}}{\text{standard deviation of noise floor}} = 
\frac{\sigma_p}{\sigma_n}
\label{eq:SNR}
\end{equation}

\section{RESULTS \& DISCUSSION}

\begin{figure}
    \centering
    \includegraphics[width=0.5\linewidth]{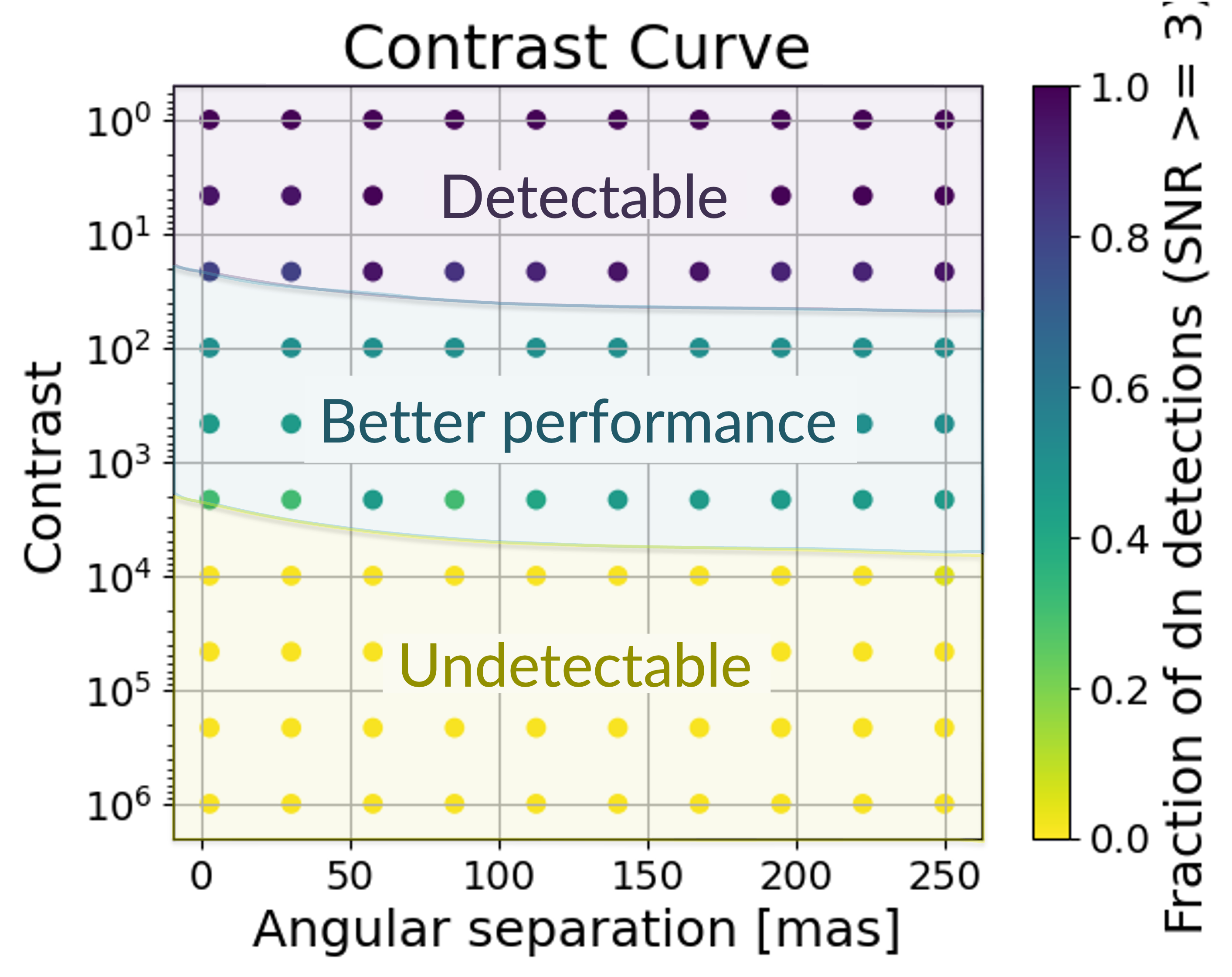}
    \caption{Preliminary contrast curve: detectability as a function of contrast and angular separation.}
    \label{fig:contrast_curve}
\end{figure}

Figure \ref{fig:contrast_curve} shows the contrast curve: maximum achievable contrast as a function of angular separation. This is after 20 seconds of observing time. Standard performance is taken as dA = 0.001, d$\phi$ = 0.01 rad. Improved performance represents a factor of 100 improvement in phase noise: dA = 0.001 and d$\phi$ = 0.0001 rad.

Simulation results show that the instrument is feasible. Maximum contrast is on the order of $10^2$, on par with current on-sky instruments. Science cases include faint companion characterization, novae imaging \cite{aydi_multiple_2026}, and exozodii characterization \cite{ertel_hosts_2018}.

Across different direct detection techniques, close-in observations are typically more difficult due to stellar photons, which reduce contrast. Our contrast curve (Fig. \ref{fig:contrast_curve}) rebukes this expectation: the curve is remarkably flat. The result shows that if a signal is detectable at any separation, it is likely visible at the innermost working angles of only a few milliarcseconds from the central source. These inner observations are complementary to on-sky coronagraphs, which typically have IWAs on the order of a few $\lambda/D$. The flat contrast curve shows how effective kernel nulling is at reducing stellar photons: because the stellar leakage is extremely low, the noise floor does not vary significantly across the instrument's field of view.

In summary, CHARM is a flexible simulation tool for the development of a nulling beam combiner for the CHARA Array. CHARM is the first simulation tool built specifically for the CHARA Array. CHARM represents the first step toward building a future H-band, self-calibrated, 6-telescope astrophotonic nulling beam combiner. Simulations show potential null depths down to contrast levels of $5\times10^2$, sufficient for observing faint companion stars, and expected to improve with increased exposure time. Instrumental performance is driven by phase aberrations, which can be improved using thermo-optic phase modulators within photonic devices. 

\section{FUTURE WORK}

Our group is currently working on a chip architecture and testbed. The first architecture will include a 2- or 4-T nulling interferometer outfitted with active phase control and several photonic taps. This chip will then be characterized using our testbed. The results of this testing will be used to iterate on our initial design. In the end, we hope to design a 6-T combiner to fully utilize CHARA's capabilities.



\bibliography{references} 
\bibliographystyle{spiebib} 

\end{document}